\documentclass[aps,prb,twocolumn]{revtex4}
\usepackage{graphicx}
\begin{document}
\title
{Present status of the theory of the high $T_c$ cuprates
}
\author{P. W. Anderson }
\affiliation{Department of Physics, Princeton University, Princeton NJ 08544}
\date{\today}
\begin{abstract}
The Gutzwiller projected mean field theory, also called Plain Vanilla or RMFT, is explained and its successes and possible extensions in describing the phenomenology of the cuprate superconductors are discussed. Throughout, we emphasize that while this is a Hartree Fock based BCS theory, it embodies fundamental differences from conventional perturbative many body theory which may be characterized by calling it a theory of the doped Mott insulator.
\end{abstract}
\maketitle
\section{Historical note}
In early 1987, just as the remarkable Bednorz-Muller discovery was becoming widely known, the basis for the theory of the materials which they had discovered was laid down \cite{pwa_87}.
It was observed that the CuO$_2$ planes on which they are based were plausibly describable by a particularly simple version of the Hubbard model, the case of a single non-degenerate band, and that the ``stoichiometric'' case where the nominal valence is Cu$^{++}$ is well described as a Mott insulator.
The superconductors are obtained when, in the "reservoir" layers between the planes, substitutional impurities of lower valence are introduced, thus doping extra holes into the Cu $d$-shell (which is of course strongly hybridized with the O $p$-shells, according to the well-known principles of ligand field theory.)  All of the plausible theories about these materials describe them as``doped Mott insulators''.

A mechanism for electron pairing in mixed valence systems, which are somewhat similar, already had been suggested by two groups \cite{pwa_book}, namely using the antiferromagnetic ``superexchange'' interaction between spins as a pairing force. In Ref. \onlinecite{pwa_87}, I likened this pairing force to the valence bonding effect for which it is essentially responsible, and pointed out that the old idea of a quantum liquid of valence bonds resonating around among different pairings of atoms had a great similarity to superconductivity.  In fact, I proposed an explicit form for such a state in terms of a Gutzwiller-projected BCS paired wave function, and in a series of papers in 1987 elaborated on formalisms for getting continuously from the Mott insulator to the superconductor.  

Unfortunately, through a series of misjudgments on my part, which are permanently recorded in an unfortunately timed book \cite{pwa_book},  my group and I thereupon fell off the correct trail to a solution, only to return to the correct path  ten years later once we had absorbed the unequivocal experimental evidence that my ``interlayer'' theory was wrong.  But fortunately, at least two separate groups had in the meantime built a theory on the 1987 foundations which turned out to be basically correct \cite{kotliar_88,zhang_88}.   In this article I will follow the second of these references but they are equivalent.  The important thing about both is that they realized that the correct solution of the original undoped RVB problem was not the isotropic ``extended $s$'' which I had been discussing but a more complex one with both $s$-like and $d$-like gaps, which Kotliar called ``$s+id$''.  Both of these papers predicted the real $d$-wave gap with nodes which was eventually observed, and in addition a number of other results which were to be confirmed one by one in the coming years.  It has been our perverse fate that the theory, properly handled, has made one after another correct prediction, well ahead of the experiments, but that these have been obscured by irrelevancies and misinterpretations until the mistaken impression has arisen that the whole subject is utterly mysterious.

It was not for another 5 years that the $d$-wave gap was verified, and by that time the field had suffered from a proliferation of proposed theories of greater or lesser degrees of plausibility.  The gradual experimental unveiling of the facts about the cuprates sometimes meant that each experiment came with a built-in theory and that theories which had predicted the result long before were not  sufficiently ``up-to-date'' to enter the public discussion.  For instance, the $d$-wave came to be identified with the idea of propagating ``antiferromagnetic spin fluctuations'', which was a popular fad at the time of its verification, rather than with its earliest, and much more natural, prediction in Ref. \onlinecite{kotliar_88,zhang_88}.  Another example of this phenomenon was the observation of the ``spin gap'' or ``pseudogap'' in underdoped materials above the superconducting ''dome'', again an obvious consequence in Ref. \onlinecite{kotliar_88,zhang_88}, but as it revealed itself it received a congeries of faddish explanations from local theorists:  a mysterious ``quantum critical point'', a ``spin nematic'', again AF spin fluctuations, the ``d-density wave'', you name it.

In any case, these early theories only came to be revived in the early 2000's by groups which were able to use them as the basis for accurate quantum Monte Carlo calculations using realistic parameter values \cite{paramekanti_02,sorella_02}, and brought forward without too much modification some of the predictions which had looked so surprising in 1988 but had been very close to correct. A group of us summarized the successes of the theory, adding a small amount of further physical ideas, in a review paper which we called the ``Plain Vanilla'' theory of high $T_c$ \cite{atoz}.  Here I will review that theory and the subsequent developments, including particularly the explanation and calculation of asymmetric tunneling spectra using it, and the recent theory of the pseudogap phase which throws a great deal of light on the overall physics of the phase diagram.

Since a great deal of emphasis has been put on the problem of the epistemics of complex phases like the high Tc cuprates, and whether a meaningful solution to the accompanying puzzles can be found , I'd like to spend a few sentences on that aspect.  
First, a bit about the nature of condensed matter physics.  Among the sciences this one is almost uniquely overdetermined, experimentally because of the variety and precision of the probes which can be applied,and theoretically because the quantum physics of atoms and electrons is so well understood.  I have always maintained that the correctness of a theoretical hypothesis is assured in this field if it can find a way to fit in with all these constraints: that there is likely to be only one possible way to fit all-or even a majority-of the observations together, and not to violate any theoretical impossibilities.  In this process of fitting things together there is no room for one-experiment theories,  doctrinal conservatism (the older generation and some younger scientists won't let go of phonons), or yet untrammeled imagination (anyon superconductivity, SO(5), QCP's, perhaps interlayer tunneling).  The naked reality is strange enough. 

A final word. The way you know you are right is when you wake up and realize that you have the answers to deep, fundamental questions that you didn't really know to ask or expect to answer.  For the old superconductors, such a question was ``why are polyelectronic metals favored?''-the question Pines, Morel, myself and McMillan answered with dynamic screening for the phonon theory.  Here there are at least two such questions: ``Why the cuprates-what is unique about copper?''; and, ``Why $d$-wave and why is the gap persistently real?  That is, why the striking nodes?'' The second is the question I didn't think to ask, but it is profound-any other simple mechanism which leads to a $d$-wave can lead also to an $xy$ or isotropic symmetry, which will appear in quadrature in order to fill in the nodes, which are intrinsically unstable in a BCS theory.  The mechanism by which the A phase of $^3$He acquires nodes was, for instance, crucial to our understanding of that system.

\section{The Plain Vanilla (RMFT) Theory}
The underlying concept of the plain vanilla theory is very simple.  In fact, it follows as closely as possible the precedent of the BCS theory.  The BCS theory in its original form is a generalization of Hartree-Fock theory to allow for not only the direct and exchange mean fields, which appear in the one-electron mean field Hamiltonian as $\bar{V}(r)\rho(r0 = \bar{v}\Psi^\ast(r)\Psi(r)$, and $A(r,r')\Psi^\ast(r')\Psi(r)$, but also the ``anomalous'' self-energy, 
$\Delta(r,r')\Psi^\ast(r)\Psi^\ast(r') + {\rm h.c.}$. These result from the three possible ways to factorize the interaction energy,
$$
\int V(r-r') \Psi^\ast(r)\Psi(r)\Psi^\ast(r')\Psi(r') dr dr'~.
$$
BCS theory is basically a variational theory:  the assumed wave function is a simple product of one-quasiparticle operators creating quasiparticles from the vacuum, and the ``gap'' equations, equivalently to the mean field equations, determine that the quasiparticle creation operators all have positive energies, so that all possible single-particle excitations increase the energy.

Simple Hartree theory won't work for a Hubbard model in which the on-site interaction energy is the largest energy in the problem.  Very early on \cite{gros_86},
it was realized that the solution to that problem was to transform to a representation in which the on-site interaction energy $U$ has been renormalized to $\infty$ as opposed to the conventional scheme well described by Shankar \cite{tsai_04} where the idea is to transform to some system of noninteracting entities. We employ a canonical transform $\exp{iS}$ to eliminate all matrix elements of the Hamiltonian which lead into the subspace in which two electrons simultaneously occupy the same site; \textit{i.e.}, those which have the large energy $U$. This transformation \cite{gros_86} can be derived perturbatively as a series in inverse powers of $U$.  That is, we start from the ``real'' Hamiltonian
\begin{equation}
H_0 = \sum_{\langle ij\rangle\sigma} t_{ij} c^\dagger_{i\sigma} c_{j\sigma} + {\rm h.c.} + U \sum_i n_{i\uparrow}n_{i\downarrow}~,
\label{hubbard}
\end{equation}
along with direct exchange and smaller terms, and transform it into the $t-J$ Hamiltonian; $H_0 \rightarrow e^{iS}H_0e^{-iS} = H_{t-J}+ O(t^3/U^2) + \dots$, where $H_{t-J}$ is given by,
\begin{equation}
H_{t-J} = \sum_{\langle ij\rangle\sigma} t_{ij} P_G c^\dagger_{i\sigma} c_{j\sigma} P_G + {\rm h.c.} + \sum_{\langle ij\rangle} J_{ij} (\vec{S}_i \cdot \vec{S}_j - \frac{1}{4} n_in_j)~.
\label{tj}
\end{equation}
In the above equation, we have ignored terms including longer-range Coulomb and phonon interactions-which latter are not particularly small, but clearly are incapable of causing the gigantic superconducting gaps which are observed. Here, $P_G$ is the full Gutzwiller projector which hereafter we will call $P$:
\begin{equation}
P = \prod_i (1-n_{i\uparrow}n_{i\downarrow})~.
\label{P}
\end{equation}
That (\ref{tj}) is really a correct description of the electronics of the cuprates was tested first by Schluter \textit{et al.} \cite{hybertsen_89}, in 1988, who found that the calculated energies of low-lying states in small clusters of the cuprate structure, using the full Hamiltonian, were well reproduced by the truncations implicit in equations (\ref{hubbard}) and (\ref{tj}).  (Another early discovery long since forgotten.)

Always remembering that the $t-J$ model wave function must be transformed by $\exp{iS}$ in the end to represent Hubbard model reality, we proceed to try to find a variational ground state for (\ref{tj}).  Clearly, since the Hamiltonian is now in block diagonal form, any low-energy state must contain only amplitudes for the projected subspace, so that,
\begin{equation}
\Psi = P \Phi(r_1,r_2,\ldots ,r_N)~,
\label{ansatz}
\end{equation}
where $\Phi$ is a general $N$-particle wave function. The essence of the ``plain vanilla'' approximation is to propose that we approximate $\Phi$, 
the wave function to be projected, using the Hartree-Fock-BCS ansatz that it is a product of quasiparticles.  I can see no reason that this is apriori less reasonable than the BCS theory itself.  If there is a single-particle-like representation of the ground state, this is the way to derive one.  In the event, there is such a representation, experimentally-by now there is all kinds of evidence that the state has gapped quasiparticles near a large Fermi surface, over a fairly wide range of doping-the so-called ``dome'' region of the phase diagram of $T$ vs doping.  I can't too much emphasize this: this procedure is the natural, and probably the only, way to derive a BCS-like superconductor from the $t-J$ Hamiltonian.

A second, and less certain, fact is that the resulting excitations may be reasonably sharp and well-defined-though, because of the projection operator, the same may not be said of actual quasiparticles : $cP$ is not the same as the single-particle-like excitation $Pc$.  But the representation in terms of $Pc$'s has some as yet unresolved peculiarities: it is overcomplete, which may mean, among other things, that the excitations can scatter each other very strongly.  But the fact of overcompleteness does not much affect either the variational equations nor the validity of them as giving the energies of approximate single-particle excitations.  
In writing out these equations we follow Ref.\ \onlinecite{zhang_88} in self-consistently choosing a particular relative gauge \cite{pwa_05} for the $J$-term relative to the kinetic energy. This choice is discussed later.

Our Ansatz for $\Phi$ in Eq. (\ref{ansatz}) ,then, is
\begin{equation}
\Phi = \prod_k (u_k+v_ k c^\dagger_{k\uparrow}c^\dagger_{-k\downarrow})
|0\rangle~.
\label{bcs}
\end{equation}         
In the recent papers by Paramekanti \textit{et al.}, the parameters $u$ and $v$ were evaluated variationally using Variational Monte Carlo techniques \cite{paramekanti_02}. But the results were almost identical to those found in the earlier papers using a very simple approximation due to Gutzwiller, which is exact in the limit that the gap is small relative to the Fermi energy.  In this approximation we assume that the correction to the probability of occupation of the sites caused by projection is uncorrelated spatially, because, obviously, the projection operates only site by site, ignoring the occupancy of neighbors. Thus the correction may be estimated by simply calculating what  happens to the average occupancies.  It is easily shown that the change in the average number of neighbors with one site empty, the other singly-occupied, is a reduction by the factor $g=2x/(1+x)$, while the change in the number of pairs of singly-occupied sites is an increase by $g_J = 4/(1+x)^2 = (2-g)^2$.
Thus the effect of the kinetic energy is reduced by the factor $g$, and that of $J$ is increased by $gJ$ , but otherwise, in this approximation, we employ the $t-J$ Hamiltonian (\ref{tj}) in precisely the same way as a real one. Thus we arrive at the ``plain vanilla'' gap equations in the ``Gutzwiller approximation'', \textit{i.e.}, the Renormalized Mean Field Theory:
\begin{eqnarray}
\Delta_k &= & g_J J \sum_{k'} \gamma_{k-k'} \frac {\Delta_{k'}}{2E_{k'}}\nonumber \\
E^2 _k &= & \xi^2_k + \Delta_k^2 \nonumber \\
\xi_k = g \epsilon_k + \varsigma_k &= & g \epsilon_k +
g_J J \sum_{k'} \gamma_{k-k'} \frac {\xi_{k'}}{2E_{k'}} \nonumber \\
\label{gapeqn}
\end{eqnarray}
Here, $J\gamma_{k-k'}$ is the Fourier transform of the exchange interaction (assumed nearest neighbor)
$\epsilon_k$ is the bare, unrenormalized kinetic energy, $\Delta$ and $\zeta$ are the anomalous and normal self energies, $\xi$, the renormalized kinetic energy and $E_k$ is the quasiparticle energy.

\begin{figure}[b]
  \centering
 \includegraphics*[width=0.45\textwidth]{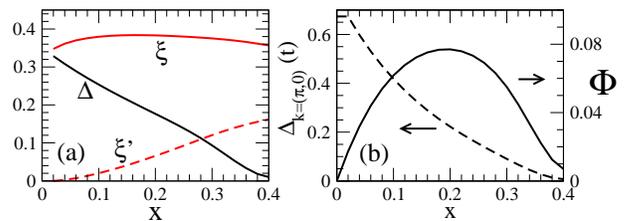}
 \caption{(a) Doping dependence of the dimensionless
mean field parameters $\xi$, $\xi'$, $\Delta$; (b) Doping
dependence of (solid) the SC order parameter, $\Phi$, and (dashed)
the gap, $|\Delta_k|$, at $k=(\pi,0)$ in units of $t$.}
\label{orderpara}
\end{figure}

\begin{figure}[b]
 \centering
 \includegraphics*[width=0.4\textwidth]{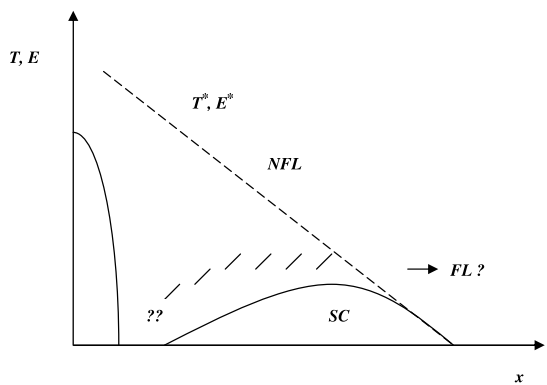}
 \caption{Generalized phase diagram of the cuprate superconductors}
 \label{phasediag}
\end{figure}

In Fig.\ \ref{orderpara}, we present results for the magnitude of the $d$-wave gap, $\Delta$, and the size of the order parameter from Edegger \textit{et al.}'s solutions of the gap equations \cite{edegger_05}, just to convince the reader that these track the observed maximum gap and dome reasonably well.
A generalized phase diagram  incorporating the results of a number of experiments is shown in Fig.\ \ref{phasediag}  (This figure differs from a phase diagram often drawn for which the $T^\ast$ line intersects the dome and no trace of the pseudogap phase remains for optimally doped materials.  Ong's Nernst effect data among others seem to unequivocally reject this interpretation.)  Since 1988, it seems, the quantitative explanation of high $T_c$ superconductivity has been available.
\section{Extensions of the RMFT}
\subsection{Spin-Charge Locking}
Note that as $g\rightarrow 0$  (the ``true'' RVB), $\Delta$ and $\varsigma$ are interchangeable. This represents a deep reality: that for the half-filled Mott insulator, the representation of the magnetic state of the spins by fermionic variables - the ``spinons'' of RVB theory-is doubly overcomplete . One may represent an $\uparrow$ spin on site $i$ either by creating an $\uparrow$ spin, $c^\dagger_{i\uparrow}$ , or by destroying a $\downarrow$ on that site, $c_{i\downarrow}$, or by any unitary superposition of the two.  In terms of  a hypothecated RVB state, described as a Gutzwiller projected BCS wave function at half filling, this means that the three Anderson-Nambu spinors $\tau_i$ ($i=1,2,3$) of the BCS state may be rotated at will, since they represent quantities which transform into each other when the SU(2) transformation is applied.  The constraint of the Gutzwiller projection also requires that only two of the three $\tau$ vectors have finite self-energies attached to them, so that  the symmetry is fully expressed as  local rotation of a dyad of self-energies $\Delta$ and $\zeta$ which must be perpendicular to each other.  All of the various alternative states which have been proposed-the ``flux phase'', the $d$-density wave, the staggered flux phase, \textit{etc.}, are one or another of these totally equivalent states, in the half-filled case.
The two ``gaps'', for the minimum-energy solution of Eq. (\ref{gapeqn}), are of maximally different symmetries.  In the half-filled case, and in the special case that we have only nearest neighbor exchange so that $\gamma$ is of the form $\cos k_x + \cos k_y$ , the two are equal in magnitude and of the form $\cos k_x \pm \cos k_y$. The only point where both vanish is where both $k$'s are $\pi/2$, which gives the nodes which are the common feature of all the equivalent ``ghost states'' I mentioned above.

It is irrelevant that the actual half-filled band is not the RVB state but a commensurate antiferromagnet, which has slightly lower energy for the Heisenberg model. It is still meaningful to examine the solutions of the full gap equation by referring them back to the hypothetical limit  $g=0$.  What happens is that, as we reintroduce the kinetic energy by doping, the antiferromagnetic state does not gain  kinetic energy as rapidly as the best RVB state, and the latter prevails at a few percent doping.  Actually, the equations (\ref{gapeqn}) represent a special choice of gauge, and we could in principle orient the kinetic energy along any chosen axis in the  $\tau$-space, and minimize the energy as a functional of that orientation-the resulting equations are given elsewhere \cite{pwa_05}. But it is clear that the optimum kinetic energy is achieved when the ``$\zeta$'' axis, the function with the symmetry $\cos k_x + \cos k_y$, is chosen as an ordinary self-energy as in (\ref{gapeqn}). 
Then the other form of solution, the odd combination $\cos k_x - \cos k_y$, acts in the direction $\tau_1$ and serves as an anomalous self-energy or gap function.  This is the principle I called ``charge-spin locking'' \cite{pwa_05}.  The locking energy was estimated in that reference as well as by Kotliar and Liu and found to be large: of order $gt$ for small dopings and comparable with $T^\ast$ for larger ones. 

This large locking energy means that the gap structure is established at temperatures well above the superconducting ``dome'' of $T_c$'s.  The reason the system does not become superconducting is that the phase stiffness is weaker, at least for doping up to the optimum, than the gap energy, in contrast to the BCS case.  
$T_c$ is determined by the proliferation of vortices, not by the breakdown of pairing.  Experimentally, in systems which are basically two-dimensional, one sees Kosterlitz-Thouless transitions;  and the cleanest measurements for optimal YBCO find 3D $XY$ model exponents, very accurately \cite{pasler_98}. Both observations indicate that the order parameter amplitude remains finite above $T_c$, and in fact the observations of Ong on Nernst effect and nonlinear diamagnetic susceptibility
\cite{npo} show that a vortex liquid state persists well above the dome, especially on the underdoped side.  From these measurements, as well as theory, we are beginning to establish that what has been called the "vortex liquid", \textit{i.e.}, a disordered superconductor as opposed to a normal metal, may be a distinct state of matter which is particularly characteristic of the cuprates.  

\begin{figure}[b]
\includegraphics*[width=0.4\textwidth]{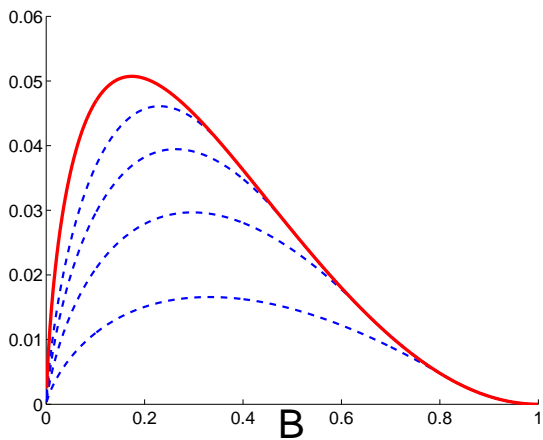}
\caption{A model calculation for the Nernst observations.}
\label{mag}
\end{figure}

That $T_c$ embodies a transition to a vortex liquid state suggests a phenomenology of this metallic state above $T_c$ quite different from that of a normal metal.  We must think of it as everywhere superconducting, but filled with a tangle of thermally-generated vortices (at low fields.) The supercurrent is fluctuating arbitrarily and the state is characterized by a persistence time $\tau$ for the supercurrents: $\langle J(0) J(t) \rangle = \langle J^2 \rangle \exp{-t/\tau}$. One may estimate that  $\tau$ is self-generated by the vortices themselves and is of order $h/\tau =h^2n_V/m$.  The conductivity of such a vortex tangle will be $\sigma = \rho_S T$.
We may speculate that when $h/\tau$  drops below $kT$, or equivalently when the number of vortices drops below a critical value where their entropy no longer compensates for their kinetic energy, the vortices evaporate: this is $T_c$, described in a Kosterlitz-Thouless way as suggested by Lee \cite{lee_05}. This provides a basis for the empirical rule proposed by Homes \cite{homes_04}, as well as for the observations of Timusk on anomalous increase of $\tau$ in the pseudogap region
\cite{puchkov_96}. An even more speculative argument based on the vortex tangle can explain the Nernst observations. (Fig.\ \ref{mag} shows a heuristic first attempt at a description of the Nernst observations). The fact that $T_c$ is controlled by the vortex liquid transition invalidates most intuitions about it from BCS theory-for instance, it makes the $d$-wave $T_c$ insensitive to scattering.
It is from the locking principle that the two insights mentioned in the Introduction arise.  Why the CuO$_2$ planes?  Because they have the feature that  nearest-neighbor exchange with only four neighbors allows the two almost degenerate gap functions of even and odd symmetry in $x$ and $y$, of which one may be used to enhance the kinetic energy, the remaining one giving a strong $x^2-y^2$ pairing energy.  Of course, there are other aspects, particularly the Jahn-Teller distortion which enhances the energy scale, and the fact that Cu$^{+++}$ does not self-trap, and all mean that unfortunately the scenario is unlikely to be repeated.   Why the nodes?  Because the RVB can only be a dyad: the spin interaction does not have a third possibility for pairing.  Thus only one function can be left over as a gap function, and it must have nodal lines which do not lie along the Fermi surface.

\subsection{Hole-Particle Asymmetry}
\begin{figure}[t]
\includegraphics*[width=0.4\textwidth]{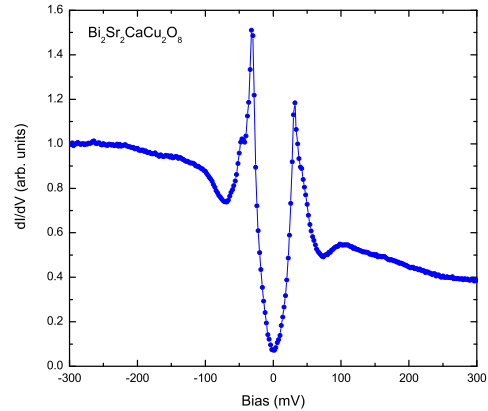}
\caption{Tunneling spectrum in optimally doped BSCCO. Data from S. H. Pan (unpublished).}
\label{tunneldata}
\end{figure}

One of the more significant experimental anomalies of the cuprates is the marked hole-particle asymmetry of the vacuum tunneling spectra.  To those of us who worked on BCS superconductivity theory, this is particularly striking because it is never observed in those materials.  There is a large ``peak-dip-hump'' structure observed on the side on which holes are injected, becoming stronger as the sample is underdoped (see Fig.\ \ref{tunneldata} ).  The underlying band structure is not responsible since it is theoretically irrelevant and experimentally implausible.  
In tunneling, a theorem of Schrieffer removes much of the effects of quasiparticle interactions, so that the broad spectra seen in ARPES are referred back to the quasiparticle pole energies; the ``hump'' structure in fact has a strong resemblance to the incoherent part of the ARPES EDC's.  It is a remarkable achievement of ``plain vanilla'' that it can give a sometimes quantitative account of these spectra.
In order to do so we must modify the  ansatz Eq. (\ref{bcs}) for $\Phi$ in Eq. (\ref{ansatz}).  BCS functions are wave-packets in the space of total electron number and one makes up non-number-conserving quasiparticles by taking advantage of this fact. This grand canonical approach is justified because the packet is centered at the correct particle number and the amplitudes for $N-2$, $N$ and $N+2$ are essentially identical.  But the projection process, while it does not change particle number, does project out very different numbers of states, so that after projection the wave packet is skewed in $N$-space.  In order to move the center of the packet back to $N$, we must introduce a fugacity factor dependent upon $N$: 
\begin{equation}
\Phi \rightarrow g^{-(n_\uparrow+n_\downarrow)/2}~,
\label{fugacity}
\end{equation}
and $g$ turns out to be the familiar kinetic energy renormalization factor $2x/(1+x)$.  Although in (\ref{fugacity}) it is clear that the factor $g$ cannot change any energy calculation since the Hamiltonian and projection conserve particle number, it is vital in understanding the process of tunneling where a particle is added or removed. Eq. (\ref{fugacity}) may be rewritten by distributing the factors of $g$ among the terms of the product, appearing very different but actually this is an obvious identity:
\begin{eqnarray}
\Phi &=& 
\prod_k
(u_k+v_ k c^\dagger_{k\uparrow}c^\dagger_{-k\downarrow})
|0\rangle \nonumber \\
& \propto& \prod_k
(\tilde{u}_k+\tilde{v}_ k c^\dagger_{k\uparrow}c^\dagger_{-k\downarrow})
|0\rangle ~,
\label{fugacitybcs}
\end{eqnarray}
where, $\tilde{u}_k = gu_k/\sqrt{g^2u^2_k+v^2_k}$ and
$\tilde{v}_k = v_k/\sqrt{g^2u^2_k+v^2_k}$.
In (\ref{fugacitybcs}), the ratio of probabilities of zero and single occupancies is correct for the projected state and is thus not altered by projection.  What it makes clear is that the projected state is made up from singlet pairs in which the relative amplitude of paired holes (the $u$ term) is decreased relative to that of paired electron spins (the $v$ term) by  the factor $g$.  In a sense, there are two types of condensed bosons, the valence bonds of the RVB and the hole pairs, and in this theory we set their relative amplitudes free, although they remain coherent:  they are ``locked'' together.
The principle on which we calculate the tunneling spectrum is the following,  Once we have chosen the form (\ref{fugacitybcs}) for $\Phi$, we may define the single-particle excitations whose energies satisfy the gap equations in terms of the wave functions, $Pc_{i\sigma}\Phi$ and $Pc^\dagger_{i\sigma}\Phi$, or equivalently, $Pc_{k\sigma}\Phi$ and $Pc^\dagger_{k\sigma}\Phi$, 
and these are now equivalently normalized.  But the matrix elements of the tunneling process insert a particle or a hole prior to the projection operation, at a particular site effectively, so that they connect to the operators $c_{i\sigma}P$ and $c^\dagger_{i\sigma}P$, and we have to commute the fermion operator through the projection operator to determine its effect. 
 
We may write  $c^\dagger = P c^\dagger + (1-P) c^\dagger$ where $(1-P)$ projects onto states with a doubly-occupied site which are effectively at infinite energy (after the canonical transformation).  Thus whenever the inserted particle encounters an occupied site,
the state is projected out, and only with probability $x$ does it encounter an empty site, \textit{i.e.}, $Pc^\dagger$.  But when it is $Pc^\dagger$ 
it lands in a legitimate excitation, \textit{i.e.}, $P c^\dagger P = P c^\dagger$.  Thus when a particle enters (with probability $x$) it does so coherently.
The hole problem is less obvious.  $c$ may be commuted through $P$ with the result,
\begin{eqnarray}
c_{i\sigma} P &=& P(1-n_{i-\sigma})c_{i\sigma} \nonumber \\
&=& P(1-\frac{1}{N}\sum_k n_{k-\sigma}-\sum_{k\neq k^\prime}c^\dagger_{k-\sigma}c_{k^\prime -\sigma})c_{i\sigma}~.
\label{cP}
\end{eqnarray}
The second term, when acting on $\Phi$, is simply a number times $c$: $\langle n_{i-\sigma}\rangle = (1-x)/2$. The third term is genuinely incoherent, creating three excitations; but these three can come from any energy in the spectrum so we expect this term to be quite small everywhere and to rise only as the square of the tunneling voltage for small voltages. The net effect of (\ref{cP}), then, is that 
\begin{equation}
cP \approx \frac{1+x}{2} Pc~.
\label{cPapprox}
\end{equation}
Thus the ratio of the probabilities of tunneling of electrons vs holes is (no surprise!) $g=2x/(1+x)$.  At high energies $\omega \gg \Delta$, where the quasiparticles are pure holes or electrons, this is the expected asymmetry, and insofar as experiment is able to ascertain, apparently this ratio agrees well (taking into account the small error caused by the canonical transformation $\exp{iS}$).

The spectrum at lower energies is complicated by the fact that superconducting quasiparticles are mixtures of electrons and holes.  At the Fermi surface, exactly at the gap energy, they are equal mixtures and the singularity at the gap must be identical for the two sides.  The working out of the exact interpolation formula for the tunnel current is a little complicated and I give here only the formulas: the tunneling density of states for electrons is
\begin{equation}
N_{e}(E,\Delta)={{d\epsilon}\over{dE}} g \left( {{u^2}\over{\sqrt
{u^2+v^2g^2}}}
+ {{v^2}\over{\sqrt{v^2+u^2g^2}}} \right).
\label{Ne}
\end{equation}
The $g$ factor in this formula comes from the projection factor-which I emphasize does not multiply the matrix element, it is essentially a relative number of open channels.  As we see, for $v \approx 1$, at high voltage, the limiting value, $1$, comes from the second factor and the tunneling is suppressed by $g$.  
On the other hand, for holes the tunneling density is
\begin{equation}
N_{h}(E,\Delta)={{d\epsilon}\over{dE}}g
\left( {{v^2}\over{\sqrt{u^2+v^2 g^2}}} +
{{u^2}\over{\sqrt{v^2+u^2 g^2}}} \right ).
\label{Nh}
\end{equation}                   
Here the $g$ factor comes from the normalized fugacity factor, and at high voltage $u \approx 1$, $v \approx 0$ and $g$ cancels out, giving the ratio $g$ between the two limits. These formulas fit data surprisingly well.

\begin{figure}[t]
\includegraphics*[width=0.4\textwidth]{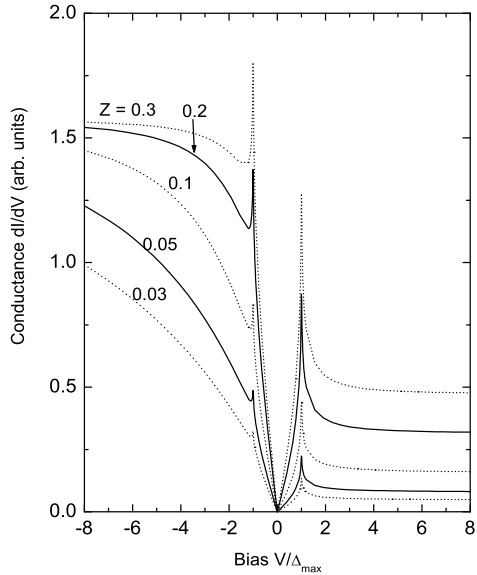}
\caption{Predicted tunneling spectra for various $Z (=g)$.}
\label{tunnelcalc}
\end{figure}

These are the formulas for fixed $\Delta$.  Note that for  $\epsilon = 0$, at the Fermi level, $u=v$ and the two are identical, the ``coherence factor'' amounting to 
$g/\sqrt{1+g^2}$. This agrees with sum rule arguments.  The asymmetry begins, however, with a vertical slope at $\Delta$, so cannot to be said to be exclusively a background phenomenon, as is seen most clearly in the fact that the peaks of observed spectra (see Fig. \ref{tunneldata}) appear to sit on background levels of different heights. These formulas must be integrated over the $d$-wave distribution of gap values to give a prediction for comparison with observed spectra.  This we have done only roughly, using $P(\Delta) = 1/\sqrt{1-\Delta^2}$,
as though the Fermi surface were circular and
not taking into account the actual band structure, which does somewhat affect the distribution of $\Delta$ values. In Fig.\ \ref{tunnelcalc}, we give the predicted spectra for a number of values of $g$, using this simplification.

The fit to experiment, at least in the main features, is fundamentally significant. Of course, it helps confirm the basic structure of the theory, and the use of superexchange as the major pairing interaction.  But it has even deeper implications. One is that even though it is basically a mean field theory based on an Hartree-Fock ansatz, it is not a Fermi liquid-based theory, that is to say that in no way can it be adiabatically continued to a BCS-like modification of Fermi liquid theory. The most fundamental property of Fermi liquid theory is hole-particle symmetry \cite{pwa_00}; after all, how can one have a theory based on a distribution of quasiparticles unless that distribution counts particles minus holes, 1 for 1? This projective theory has destroyed that symmetry in a very fundamental-yet simple-way.  

Yet the projective feature is rooted in the real physics of the system.  As pointed out also by Capello \textit{et al.}, once one is above the Mott critical $U_c$, there are what we  called ``anti-bound states'' \cite{hsu} and what Ref.\ \onlinecite{pwa_00} calls ``holon-doublon bound states'' which cannot be treated perturbatively but must simply be projected out of the problem.  A result is that the spectrum is overcomplete and the particle operators do not obey a simple fermion algebra.  I believe that the broad features seen in the momentum-resolved particle spectra are related to that problem, but fortunately the tunneling spectrum is simplified by the benign results of ``Schrieffer's theorem'' and is easier to interpret. 

\subsection{Fluctuations of the Asymmetry Parameter $g$}
The asymmetry parameter, as I remarked, plays a role similar to that of a condensate of hole pair bosons, which is locked to the spinon pairs of the RVB by the charge-spin locking process.  Its average value is determined by the doping and charge neutrality, but it is evident that formally we can allow it to vary either in space or in momentum space.  More speculatively, we can allow it a dynamic character, and I believe that its collective modes do in fact give us extra degrees of freedom which play a role in the non-Fermi liquid behavior mentioned above.  This and the other remarks I will make in this rather speculative section are inspired more by suggestions from experimental observations than by apriori theory, but I do believe that treating g as a physical object can lead to considerable insights.

The most obvious is the possibility of allowing $g$ to vary along the Fermi surface, just by making it a function of $k$ in (\ref{fugacitybcs}).  The mean value $x=\langle g/(2-g)\rangle$ must be maintained for charge neutrality.  The phenomenon of ``Fermi arcs'' is observed by ARPES in underdoped systems, where the regions of the Fermi surface near the nodes remain sharply defined while the antinodal regions smear out and disappear. The nodal regions are also those where the kinetic energy is greatest, so that one could gain energy by making $g$ large at the nodes and small at the zone corners.  No calculation of this proposed effect yet exists.

It has also been proposed that the hole percentage may vary spatially, in particular that at low doping $g$ could form a kind of charge density wave or ``superconducting electron solid'' \cite{pwa_04}.  The motivation could be Madelung energy of the pairs;  or it is possible that there is a tendency to bistability near the low-doping quantum critical point.  

Finally, there is the question of phase fluctuations of $g$ (which is the appropriate variable to assign a phase to, since it controls the charge carriers.)  It is known that the phase transition at $T_c$ is of ``$X-Y$'' character both for optimal doping \cite{pasler_98} and for very low doping near the quantum critical point \cite{franz_05} , and that above $T_c$ there is a large region in which the state is best described as a ``vortex liquid'' rather than a normal metal, \textit{i.e.}, there is a fluctuating superconducting order parameter. (see Ref.\ \onlinecite{npo}) This has been described as a regime in which $\zeta$ is still locked to the kinetic energy but $\Delta$, \textit{i.e.}, the phase of \textit{g},  is freely fluctuating \cite{pwa_05}. There is a very important open question here as to whether or not there is a transition into a still higher T phase which has an RVB but is not a vortex liquid \cite{nguyen_98}. 

\section{Discussion: Alternative approaches}
The RMFT theory works, based on $t-J$ physics and superexchange as the interaction, and can account semiquantitatively for the basic phenomena of cuprate superconductivity, and qualitatively for many more. Why then are contradictory theories being promoted? The most popular theories reject the Mott-Anderson physics entirely and go in contrary directions. There seems to be a psycho-social need among physicists for an explanatory \underline boson, some kind of tangible glue to hold the pairs together, I suppose because of the folk memory of talks about the BCS mechanism and the analogy of two bodies on a mattress; or else a simplified view of Feynman diagrams. It is felt, I suppose, that the Mott theory is based on purely repulsive forces-but those of us who actually worked on BCS recognize that the phonon interaction is not literally an attraction either, merely a partial screening of the electrons' Coulomb repulsion. Why a superexchange integral universally agreed and experimentally measured to be of order 1000 degrees is thought to be inadequate for pairing has always escaped me; but it is.  The two most popular glues are phonons and antiferromagnetic spin fluctuations.

Phonons start out with a big disadvantage: the BCS concept is irrevocably based on an on-site, \textit{local} interaction; and is incompatible with $d$-wave.  In the cuprates, the phonons are undoubtedly optical ones involving the oxygen octahedron (oh, there are other suggestions, but even less plausible) and there are perhaps ways of distorting these in order to give a $d$-wave, but I have never seen a plausible one.  Intrinsically, Einstein optical phonons lead to local interactions.  But, experiment is the best teacher. The isotope effect measurements of Keller \cite{khasanov_04} find a reasonably-sized isotope effect on $T_c$, apparently confirming the phonon hypothesis;  but Keller was thorough enough to also measure the isotope shift of $\rho_s$, the superfluid density, $\propto \lambda^{-2}$; and he finds that this shifts by the same fractional amount.  It was pointed out very early in the game by Fisher \textit{et al.}, \cite{fisher_88} that unlike the polyelectronic metals for which BCS theory works and the isotope shift comes entirely from the pairing interaction, oxides are best understood as tight-binding systems with interactions which depend exponentially on interatomic distances.  Thus zero-point vibrations will have an appreciable effect on normal state properties such as the band mass which determines $\rho_s$.  Since $T_c$ is an $X-Y$ transition as already remarked, its value is expected to be directly proportional to $\rho_s$, the coupling in the $xX-Y$ model, as observed, so that apparently there is no experimental isotope shift ascribable to the pairing interaction. In fact, even if there were, $T_c$ is insensitive to the actual value of $\Delta$, as we explained above. Extensive ARPES studies have catalogued what may be phonon effects on the quasiparticle dispersion \cite{gweon_04} but these seem to be irrelevant to the pairing mechanism.  $J$, of course, itself varies in a similar way as t with interatomic distance and may provide a partial source of the observed isotope shifts in dispersions.  It seems that calculating phonon effects, while worth doing for its own sake, is not the most urgent task.

There are other phonon schemes, most notoriously the bipolaron theory.  One understands the impulse to look this way, since polaron phenomena are so ubiquitous in oxides.  But very early on it became clear that one reason the cuprates are so favored is that this case is gloriously free of polaron effects, presumably because Cu$^{++}$ and Cu$^{+++}$ have similar Jahn-Teller displacements.  The remarkably detailed tunneling and ARPES spectra demonstrating well-characterized quasiparticles exclude small polaron phenomenology.  I believe that Baskaran's theory  \cite{baskaran_05} explaining the electron-doped case as dominated by small polarons must be essentially correct, and the contrast with hole doping illustrates well what phenomenology polarons might lead to.

A second putative source of the ``glue'' boson is ``antiferromagnetic spin fluctuations''.  This idea sounds similar to the Mott-based theory but is not at all so, in fact proceeds on exactly the opposite principle: that in the end the physics is to be obtained by ``summing all the diagrams'' starting from a Fermi liquid \cite{schmalian_97}.  Another way to say it is that the assumption is that the theory fits under the general scheme of Ref.\ \onlinecite{tsai_04}, where all interaction terms are renormalized downwards, while the plain vanilla theory makes the assumption that one must start by renormalizing $U \rightarrow \infty$, with the Rice canonical transformation.  I feel that $U_c$, the Mott critical $U$, marks a fundamental separatrix between basins of attraction, and that the cuprate case is on the large $U$ side.  The key question is whether the frequency associated with most of  the pairing interaction is above a Mott-Hubbard gap, and therefore  cannot be represented by a boson whose spectrum extends continuously to zero frequency.  In that case it might as well be represented by a simple four-Fermion vertex $J$.  The idea of antiferromagnetic spin fluctuations is that the opposite is the case, and that somehow if one can sum enough diagrams the Mott gap will disappear from the problem and interactions will proceed by the exchange of a putative low-frequency spin-fluctuation boson.

Since, in fact, one cannot come close to summing all the diagrams, papers based on this idea have tended to contain about one parameter per experimental fact, and therefore to ``explain'' great numbers of these facts.  Apparently recent advances in experimental detail have led to exhaustion of invention, and many rather crucial discoveries remain unexplained, for example Fermi arcs, tunneling asymmetry, the vortex liquid phase, the checkerboard, Homes' identity.

There are a number of more mysterious suggested sources for the ``glue boson'', many of which invoke the equally opaque concept of a hidden ''quantum critical point''; their variety excludes detailed explication.

Perhaps no longer worthy of mention is the ``stripe theory'', the problem of which was that it never seemed to be a theory of the superconductivity, but only a theory of the stripes themselves.  Since stripes are not common to many of the cuprate superconductors, and as time goes on to fewer and fewer, it is hard to understand their relevance.

\section{Conclusions and Anticipations}
Many of my conclusions were rather strongly stated in the Introduction.  It seems that the Gutzwiller method works perhaps even better than we had any right to expect.  It also has the added feature that it brings out the deep difference in principle between a Fermi-liquid based approach and the actual behavior of the cuprates in a relatively simple and straightforward way, both in demonstating the hole-particle asymmetry of the Green's functions and in the ``locking'' phenomenon.

Quite understandably, there are other ways to approach the same physical model, and some of them have a good chance of being more accurate or rigorous-for instance gauge and slave boson theories, one of which I quoted here.  One can certainly differ on the applicability of the crude approximations made in Plain Vanilla to make it soluble; and it is very meaningful to try to add in further terms to the interactions used, and to study the various accompanying phenomena such as coexistence with antiferromagnetism. The major puzzle  remains that of the Strange Metal, the mysterious phase above $T^\ast$, and the strange quantum critical point where the $d$-wave gap goes to zero.  The linear $T$, linear in   electron-electron  scattering mechanism which pervades the high-energy region is still a puzzle but must be a characteristic of the purest Mott physics \cite{fn}.

I should not fail to mention the accumulation of recent direct or semi-direct calculational results all of which are now tending to converge on the conclusion that $d$-wave superconductivity undoubtedly appears in the Hubbard and $t-J$ models.  I am sure these will be represented well elsewhere in this volume; and of course, I have absolutely no problem with them;  it is a matter of taste whether one prefers approximations such as Plain Vanilla which allow understanding of the phenomenology,  or more exact but only semitransparent calculations.

Throughout the paper I have alluded to avenues for further exploitation of the method, specifically the possible explanation of the ``Fermi arcs'' as a $k$-dependence of $g$, and of nanoscale structures as spacial modulations of it; but both will require more detailed calculations than we are yet capable of. 
\section{Acknowledgements}
I have first of all to acknowledge my long-term collaborator Nai-Phuan Ong, for the innumerable times he has helped me with experimental know-how and theoretical comments.  I should also acknowledge my collaborators in the Plain Vanilla exercise, T. M. Rice, P. A. Lee, Mohit Randeria, Nandini Trivedi and Fu-Chun Zhang, as well as others who were involved in finding the solution 17 years too early: Claudius Gros and Gabi Kotliar.  Others who have helped keep my mind clear about experimental data have been Doug Bonn, Nicole Bontemps,  Bernhard Keimer, Seamus Davis, Mike Norman, J-C Campuzano, Tom Timusk, Kam Moler;  this is only a tiny fraction of the totality of individuals who have been helpful.  But none of the above need take any responsibility for what I say here.  Finally, there is my good friend and sounding board, V Muthukumar.


\begin{thebibliography}{99}
\bibitem{pwa_87} P. W. Anderson, \textit{Valence Fluctuations and Heavy Fermions},  Gupta and Malik eds., Plenum, NY, 1987, p 9;  Science \textbf{235}, 1196 (1987); \textit{Frontiers and Borderlines in Many-Particle Physics} (Varenna Summer School 1987), Broglia and Schrieffer eds., North-Holland NY 1988, p 1.
\bibitem{pwa_book}P. W. Anderson, \textit{The Theory of Superconductivity in the High $T_c$ Cuprates}, Princeton University Press, Princeton, 
1997.
\bibitem{kotliar_88} G. Kotliar and J. Liu, Phys Rev B \textbf{38}, 5142 (1988).
\bibitem{zhang_88} F. C. Zhang, C. Gros, T. M. Rice and H. Shiba, J Supercond Sci Tech \textbf{1}, 36 (1988).
\bibitem{paramekanti_02} A. Paramekanti, M. Randeria, and N. Trivedi, Phys. Rev. Lett. \textbf{87}, 217002 (2002); cond-mat/0101121(2001).
\bibitem{sorella_02} S. Sorella, G. B. Martins, F. Becca, C. Gaza, L. Capriotti, A. Parola, and E. Dagotto, 
Phys Rev Lett. \textbf{88}, 117002 (2002); cond-mat/0110460 (2001).
\bibitem{atoz} P. W. Anderson, P. A. Lee, M. Randeria, T. M. Rice, N. Trivedi, and
F. C. Zhang, J. Phys. Cond. Mat. {\bf 16}, R755 (2004).
\bibitem{gros_86} C. Gros, R. Joynt, and T. M. Rice, Phys. Rev. B \textbf{36}, 381 (1986), and references therein.
\bibitem{tsai_04} S-W Tsai, A. H. Castro Neto, R. Shankar, and D. K. Campbell, cond-mat/0406174 (2004).
\bibitem{hybertsen_89} M. S. Hybertsen, M. Schluter, and N. E.  Christensen, Phys. Rev. B \textbf{39}, 9028 (1989).
\bibitem{pwa_05} P. W. Anderson, cond-mat/0504453; to appear in \textit{Phys. Rev. Lett.}
\bibitem{edegger_05} Bernhard Edegger, V. N. Muthukumar, Claudius Gros, and P. W. Anderson, \textit{to be published}.
\bibitem{baskaran_88} G. Baskaran and P. W. Anderson, Phys. Rev. B \textbf{37}, 580 (1988).
\bibitem{pasler_98} V. Pasler \textit{et al.}, Phys. Rev. Lett. \textbf{81}, 1094 (1998).
\bibitem{npo} N. P. Ong and Y. Y. Wang, cond-mat/0306399; N. P. Ong, \textit{et al.}, Ann. der Physik \textbf{13}, 9 (2004); N. P. Ong, and Wei-Li Lee, cond-mat/0508236; Y. Wang \textit{et al.}, cond-mat/0503190.
\bibitem{lee_05} P. A. Lee and X. G. Wen, \textit{to appear in Rev. Mod. Phys.}
\bibitem{homes_04} C. C. Homes \textit{et al.}, cond-mat/0410719; Nature \textbf{403}, 746 (2000).
\bibitem{puchkov_96} A. V. Puchkov, D. N. Basov and T. Timusk, J Phys Cond Matt \textbf{8}, 10049 (1996).
\bibitem{pan_00} S. H. Pan \textit{et al.}, Nature \textbf{403}, 746 (2000).
\bibitem{pwa_tunnel} P. W. Anderson and N. P. Ong, cond-mat/0405518 ;  J. Phys. Chem. Solids, \textit{to be published}.
\bibitem{pwa_00} P. W. Anderson and F. D. M.  Haldane, cond-mat/0009309 (2000); J Stat Phys \textbf{103}, 425 (2001).
\bibitem{capello_05} M. Capello, F. Becca,, M. Fabrizio, S. Sorella, and E. Tosatti, Phys. Rev. Lett. \textbf{94}, 026406 (2005).
\bibitem{hsu} T. C. Hsu, Ph. D. Thesis, Princeton University (1989).
\bibitem{pwa_04} P. W. Anderson, cond-mat/0406038.
\bibitem{franz_05} M. Franz and A. Iyengar, cond-mat/0504735.
\bibitem{nguyen_98} A. K. Nguyen and A. Sudbo, cond-mat/9811149; Europhys. Lett. \textbf{46}, 780 (1999).
\bibitem{khasanov_04} R. Khasanov et al., Phys. Rev. Lett. \textbf{92}, 057602 (2004).
\bibitem{fisher_88} D. S. Fisher, A. J. Millis, B. Shraiman, and R. N. Bhatt, Phys. Rev. Lett. \textbf{61}, 482 (1988).
\bibitem{gweon_04} G. H. Gweon, S. Y. Zhou, and A. Lanzara, J. Phys. Chem. Solids \textbf{65}, 1397 (2004); cond-mat/0404720; cond-mat/0407556.
\bibitem{baskaran_05} G. Baskaran, cond-mat/0505509.
\bibitem{schmalian_97} J\"org Schmalian, \textit{informal talk}, 1997.
\bibitem{fn} The Strange Metal problem seems to have been solved, in  very recent work which has not yet been written up. The phenomena turn out to be surprisingly direct consequences of Gutzwiller projection.
\end{thebibliography}
\end{document}